\documentclass[prb, preprint]{revtex4}
\usepackage{graphicx}
\usepackage{epsfig}
\usepackage{rotating}

\begin{document}

\title{Side-chain and backbone ordering in Homopolymers}

\author{Yanjie Wei} 
\email{yawei@mtu.edu}
\affiliation{ Department of Physics, Michigan Technological University, 
             Houghton, MI 49931, USA}

\author{Walter Nadler}
\email{wnadler@mtu.edu}
\affiliation{Department of Physics, Michigan Technological University, 
             Houghton, MI 49931, USA}

\author{Ulrich H.E. Hansmann} 
\email{hansmann@mtu.edu, u.hansmann@fz-juelich.de}
\affiliation{Department of Physics, Michigan Technological University, 
             Houghton, MI 49931, USA}
\affiliation{John-von-Neumann Institute for Computing, 
             Forschungszentrum J\"ulich, D-52425 J\"ulich, Germany}

\date{\today}

\begin{abstract}
In order to study the relation between backbone and side chain ordering in proteins, we have performed multicanonical simulations of deka-peptide chains with various side groups. 
{\it Glu$_{10}$, Gln$_{10}$, Asp$_{10}$, 
Asn$_{10}$, and Lys$_{10}$} were selected to cover a wide variety of possible interactions between the side chains of the monomers. 
All homopolymers undergo helix-coil transitions.
We found that peptides with $long$ side chains that are capable of hydrogen bonding, i.e. {\it Glu$_{10}$,} and {\it Gln$_{10}$}, exhibit a second transition at lower temperatures connected with side chain ordering. This occurs in gas phase as well as in solvent, although the character of the side chain structure is different in each case. However, in polymers with $short$ side chains capable of hydrogen bonding, i.e. {\it Asp$_{10}$} and {\it Asn$_{10}$}, side chain ordering takes place over a wide temperature range and exhibits $no$ phase transition like character. Moreover, non-backbone hydrogen bonds show enhanced formation and fluctuations already at the helix-coil transition temperature, indicating competition between side chain and backbone hydrogen bond formation. Again, these results are qualitatively independent of the environment.
Side chain ordering in {\it Lys$_{10}$}, whose side groups are long and polar, also takes place over a wide temperature range and exhibits no phase transition like character in both environments.
Reasons for the observed chain length threshold and
consequences from these results for protein folding are discussed.
\end{abstract}


\maketitle

\section{Introduction}

In the last two decades,  energy landscape  and folding funnel
paradigm \cite{Onuchic} have led to an emerging understanding of the
folding process in proteins. However, these concepts describe only the
general characteristics of folding. Many details still remain poorly
understood. One aspect is the role of side chain ordering in the folding
process. Are side chain and backbone ordering coupled? Do the processes
occur in a particular order? In principle, these questions can be
investigated {\it in silico}. As one cannot employ lattice proteins and
other minimalist models that ignore side chains, one has to rely on
all-atom models of proteins. However, simulations of such protein
models are often hampered by poor convergence \cite{H05}, and in particular slow side chain dynamics can pose problems \cite{Shimada2001}. Only with the
development of generalized ensemble techniques \cite{HO98} such as
parallel tempering \cite{PT1,PT2,H97f} or  multicanonical sampling
\cite{MU,MU2} have all-atom simulations of small proteins (with  up to
$\approx 50$ residues\cite{KH05}) become possible. In the present  work 
we employed multicanonical sampling \cite{MU} which was first introduced to
protein science in Ref.~\onlinecite{HO}.

In a previous publication \cite{Wei2006} we reported results on backbone and side chain ordering in polyglutamic acid, {\it Glu$_{10}$}. 
Our results showed that - upon continuously lowering the temperature - {\it Glu$_{10}$} "folding" is a two-step process, starting  with the secondary structure formation. Only after the backbone geometry is fixed the side chains order themselves at a much lower temperature. This scenario applied to the polymer in gas phase as well as in a solvent. It is quite remarkable that the scenario is independent of the particular environment since the side chain ordering had a different character in each case. In gas phase, side chains align themselves along the helical cylinder,  stabilizing themselves by forming hydrogen bonds with each other, while in solvent the side chains extend into the solvent which screens them from forming hydrogen bonds. 

These findings for {\it Glu$_{10}$}, decoupling of backbone and side chain ordering independent of the environment, immediatly raise the question whether such a scenario is a common characteristic in protein folding, at least for hydrogen bond forming side chains.

In order to test this hypothesis we are investigating in this contribution the dependence of side chain ordering on chain size and chemical properties. Glutamine (Gln), for example,  is about the size of glutamic acid and is also able to participate in hydrogen bonds. Aspartic acid (Asp) and Asparagine (Asn) are also able to form hydrogen bonds, but have a smaller size, while Lysine (Lys) has a larger polar side chain.
Consequently, we have performed multicanonical simulations of {\it Gln$_{10}$, Asp$_{10}$, Asn$_{10}$, and Lys$_{10}$}, and will compare them to extended results for {\it Glu$_{10}$}.

Again, the folding scenarios found are independent of the environment, despite the fact that the side chain structure that arises is different for gas phase and in solvent. As an aside we note that the structural features found here confirm early investigations performed by H. Scheraga's group some 40 years ago \cite{Scheraga66,Scheraga67,Scheraga68,Scheraga70}.

However, we observe a decoupling of backbone and side chain ordering transitions - as described above - only for {\it Glu$_{10}$} and {\it Gln$_{10}$}, i.e. for long hydrogen bond forming side chains. Polymers with short hydrogen bond forming side chains, i.e. {\it Asp$_{10}$} and {\it Asn$_{10}$}, exhibit no such decoupling. Instead, side chains already participate at the helix coil transition as competitors for the backbone hydrogen bonds. Furthermore, the final side chain ordering in these polymers takes place over a wide temperature range and exhibits no phase transition like character. {\it Lys$_{10}$} whose side chain is long and polar follows the same szenario as {\it Asp$_{10}$} and {\it Asn$_{10}$}, albeit without the side chains competing at the helix-coil transition.

\section{Methods}
Our simulations utilize  the ECEPP/3 force field \cite{EC} as
implemented in the 2005 version of the program package SMMP \cite{SMMP,SMMP05}.
 Here the interactions between the atoms within the homopolymer chain are approximated
 by a sum $E_{ECEPP/3}$ consisting of electrostatic energy $E_C$, a  Lennard-Jones term $E_{LJ}$,
 hydrogen-bonding term $E_{HB}$ and a torsion energy $E_{Tor}$:
 \begin{eqnarray}
  E_{\text{ECEPP/3}} &=& E_C + E_{LJ}  + E_{HB} + E_{Tor} \nonumber \\
  &=&  \sum_{(i,j)} \frac{332 q_i q_j}{\epsilon r_{ij}} \nonumber \\
 &&   + \sum_{(i,j)} \left( \frac{A_{ij}}{r_{ij}^{12}} - \frac{B_{ij}}{r_{ij}^6} \right) \nonumber \\
 &&   + \sum_{(i,j)} \left( \frac{C_{ij}}{r_{ij}^{12}} - \frac{D_{ij}}{r_{ij}^{10}} \right) \nonumber \\
  && + \sum_l U_l ( 1\pm \cos(n_l \xi_l)) \;,
\label{energy}
\end{eqnarray}
where $r_{ij}$ is the distance between the atoms $i$ and $j$,  $\xi_l$ is the $l$-th torsion
angle, and energies are measured in Kcal/mol.
The protein-solvent interactions are approximated by a solvent accessible surface term
\begin{equation}
  E_{solv} = \sum_i \sigma_i A_i \;.
\label{solventEnergy}
\end{equation}
The sum goes over the  solvent accessible areas $A_i$ of all atoms $i$ weighted by solvation
parameters $\sigma_i$ as determined in Ref.~\onlinecite{OONS}, a common choice when the
ECEPP/3 force field is utilized.
Our previous experiences \cite{PTH,TTH} have shown that $E_{solv}$ reproduces the  effects of protein-water interaction  {\it qualitatively} correct.
However, thetemperature scale is often distorted, leading, for instance, to transitions at temperatures where in nature water would be vaporized. This problem can be remedied, however, by renormalization of the temperature scale upon comparison with experiments.

The above defined energy function leads  to a landscape that is characterized by a multitude of minima separated by high barriers. As the probability to cross an energy barrier of height $\Delta E$ is given by$\exp (-\Delta E/k_BT)$, $k_B$ being the Boltzmann constant, it follows that  extremely long runs are necessary to obtain sufficient statistics in regular canonical simulations at low temperatures.Hence, in order to enhance sampling we rely on the multicanonical approach \cite{MU,MU2} asdescribed in Ref.~\onlinecite{HO}.  Here, configurations are weighted with a term $w_{MU} (E)$ determined iteratively such that the probability distribution obeys\begin{equation}    P_{MU}(E) \propto n(E) w_{MU}(E) \approx const~,\end{equation}where $n(E)$ is the spectral density of the system. Thermodynamic averages of an observable $<O>$ at temperature $T$ are obtained by re-weighting \cite{FS}:\begin{equation} <O>(T) = \frac{\int dx \; O(x) e^{-E(x)/k_BT} / w_{MU}[E(x)]}                          {\int dx \; e^{-E(x)/k_BT} / w_{MU}[E(x)]}\end{equation}where $x$ counts the configurations of the system. 

After determining the multicanonical weights $w_{MU}(E)$ we have performed  multicanonicalsimulations of $5\times 10^6$ sweeps. Each sweep consists of $N_f$ Metropolis steps that try to updateeach of the $N_f$ dihedral angles (the degrees of freedom in our system) once. Here, $N_f=70,70,60,60,80$ for {\it Glu$_{10}$, Gln$_{10}$, Asp$_{10}$, Asn$_{10}$, and Lys$_{10}$}, respectively. Every 10 sweeps various quantities are measured and written to a file for further analysis. These includethe energy $E$ with its respective contributions from Eq.~(\ref{energy}) and - in the case of the simulations in solvent - from the protein-solvent interaction energy $E_{solv}$. The radius of gyration $r_{gy}$ as a measure of the geometrical size, and the number of helical residues $n_H$, i.e. residues where the pair of dihedral angles $(\phi,\psi)$ takes  values in the range ($ -70^\circ \pm 30^\circ $, $ -37^\circ \pm 30^\circ $) \cite{Okamoto1995}.
Finally, we monitor the total number of hydrogen bonds, $n_{HB}$. Note that hydrogen bonds along the backbone span four residues in the sequence. We therefore monitor, in addition, the number of hydrogen bonds between residues that are closer in the sequence, denoted by $n_{HB}^S$.
This number is a lower limit to the number of non-backbone hydrogen bonds\cite{sidechainHBFootnote}.

\section{Results and Discussions}

Our aim is to study through multicanonical sampling 
the relationship between side chain ordering and other transitions for
five specifically selected  homopolymers. We are interested particularly in the
behavior of these molecules at low temperatures where their energies
are low, too. Hence, a measure for the quality of our results is the number of
{\it independent} low-energy states. As a multicanonical simulation performs a
random walk in energy space, see Fig.~\ref{glu_vac}, a lower limit for this number of independent states can be obtained from
the number $n_T$ of  passages from a suitably defined low-energy state to high-energies, and back, sometimes also called "tunneling processes". Since we wanted to obtain information about transitions that occur in the temperature range $200K$ to $600K$, we choose the average energies at $T=150K$ and at $T=750K$ as reference energies in each case. These reference energies and the numbers of tunneling processes for each simulation are reported in Table~\ref{tab:table1}.
As in previous investigations, we found $n_T>100$ to be sufficent for a good quality of the results. As a check we performed a $10^7$ sweep simulation of {\it Glu$_{10}$} in vacuum while all other simulations were $5\times 10^7$ sweeps.

\begin{table}
\caption{\label{tab:table1} Number of tunneling processes, $n_T$, between 
the average energy at $T=150K$, $E_{150K}$, and the average energy at $T=750K$,  $E_{150K}$
in each simulation ($5\times 10^6$ sweeps). The number in parenthesis is the error in the last digit.}
\begin{ruledtabular}
\begin{tabular}{lrrrrrr}
 & & vacuum & & & solvent & \\
 & $n_T$ & $E_{150K}$ [Kcal/mol] & $E_{750K}$ [Kcal/mol] & $n_T$ & $E_{150K}$ [Kcal/mol] & $E_{750K}$ [Kcal/mol] \\
\hline
{\it Glu$_{10}$}\footnotemark[1]  
&  303 & -132.4(2)  & -42.6(1) & 183  & -210.08(5)  & -124.61(8) \\
{\it Gln$_{10}$ } &  186 & -161.3(3)  & -76.29(8) & 251 & -232.42(5)  & -154.10(4) \\
{\it Lys$_{10}$ } &  302 & -33.6(1)   &  55.53(4) & 149 & -100.23(5)  & -10.24(6) \\
{\it Asn$_{10}$ } &  168 & -195.65(10)  & -121.14(5) & 215  & -267.52(4) & -196.86(4) \\
{\it Asp$_{10}$ } &  118 & -169.7(1)  & -90.8(1) & 146 & -243.2(9)  & -169.6(9) \\
\end{tabular}
\end{ruledtabular}
\footnotetext[1]{The {\it Glu$_{10}$} simulation in vacuum was $10^7$ sweeps for testing reasons.}
\end{table}

We first investigate the case of molecules in gas phase. Fig.~\ref{C_gas} displays the specific heat per molecule,
\begin{equation}
     C(T) = k_B \beta^2 (<E^2> - <E>^2) \quad ,
\label{specHeat}
\end{equation}
as a function of temperature for all five polymers. In each case one observes a peak at a temperature $T_1$ in the range $450K$ to $600K$ with quite large  half-widths  between $150K$ and $200K$. The individual peak temperatures $T_1$ and other properties are listed in Table~\ref{tab:table2}.

The corresponding plot of the helicity in Fig.~\ref{helicity_gas} shows that these peaks separate a high temperature region where the backbone has no ordering from a region where temperatures are low enough to allow the formation of an $\alpha$-helix.
As one can see from the monotonic drop in the average radius of gyration $r_{gy}$  that is shown in the inlay, the  helix-coil transition is also connected with a collapse of the molecule. Below the transition $r_{gy}$ stabilizes, reflecting the stable helical structure that has been reached.

The transition temperatures are largest for {\it Glu$_{10}$} and {\it Gln$_{10}$}, and lowest for {\it Asn$_{10}$} and {\it Asp$_{10}$}, the individual values being very close to each other in each block. The transition temperature for {\it Lys$_{10}$} is intermediate between both blocks. Since the side chains of {\it Glu$_{10}$} and {\it Gln$_{10}$} are larger than in the cases of {\it Asn$_{10}$} and {\it Asp$_{10}$}, they provide sterical hindrances to backbone conformations, leading to a decrease of the backbone entropy. As the transition is driven by entropy, and the energetic properties of these four heteropolmers are similar, this leads to a higher transition temperature for the polymers with larger side chains.
This argument also explains the relationship of the transition temperature for 
{\it Lys$_{10}$} to that of a homopolymer of similar energetics but much shorter side chain, {\it Ala$_{10}$} with a transition temperature $T_1=427K$, see Ref.~\onlinecite{Hansmann1999} and Table~\ref{tab:table2}.

\begin{table}
\caption{\label{tab:table2} Properties of the helix-coil transitions observed for the five homopolymers: Transition temperature $T_1$, half width $\Delta T$, and specific heat per molecule at the transition temperature, $C(T_1)$. The number in parenthesis is the error in the last digit. }
\begin{ruledtabular}
\begin{tabular}{lrrrrrr}
 & & vacuum & & & solvent & \\
 & $T_1$[K] & $\Delta T$\footnotemark[1][K]  & $C(T_1)$ [Kcal/mol] &
   $T_1$[K] & $\Delta T$\footnotemark[1][K]  & $C(T_1)$ [Kcal/mol] \\
\hline
{\it Glu$_{10}$ } &  587(14) &  161  & 0.193(2) & 477(7) &  88  &  0.296(3)  \\
{\it Gln$_{10}$ } &  584(14) &  163  & 0.181(1) & 484(8) &  97  &  0.268(3)  \\
{\it Lys$_{10}$ } &  538(8)  &  151  & 0.216(2) & 447(10) &  98  &  0.266(3)  \\
{\it Asn$_{10}$ } &  485(19) &  193  & 0.169(2) & 424(9) & 105 & 0.249(4)   \\
{\it Asp$_{10}$ } &  471(19) &  170  & 0.182(2) & 415(6) & 82 &  0.300(3)   \\
\hline
{\it Ala$_{10}$\footnotemark[2] } &  427(7) &  146  &   & 333(2) &  &     \\
\end{tabular}
\end{ruledtabular}
\footnotetext[1]{determined at $C=\left[C(T_1)+C(T_{min})\right]/2$, with $T_{min}$ from either $C'(T_{min})=0$ or $C''(T_{min})=0$.}
\footnotetext[2]{The data for {\it Ala$_{10}$} are from Refs.~\onlinecite{Hansmann1999, Peng2003} and are included for comparison.}
\end{table}

However, this entropic argument does not suffice to understand why the transition temperature of {\it Lys$_{10}$} is still smaller than that of {\it Glu$_{10}$} and {\it Gln$_{10}$}, although {\it Lys$_{10}$} has even longer side chains than those two molecules. In order to explain this feature we have to take the different energetics into account, i.e. the fact that the side chains of {\it Glu$_{10}$} and {\it Gln$_{10}$} can participate in hydrogen bonds, which are stronger than it would be for the case of { \it Lys$_{10}$}. These additional energy contributions shift the transition temperature of {\it Glu$_{10}$} and {\it Gln$_{10}$} to still higher values. This energy contribution also explains why the transition temperatures of {\it Asn$_{10}$} and {\it Asp$_{10}$} are still higher than that of {\it Ala$_{10}$}, although the side chains for those three molecules are of similar size.

The inlay of Fig.~\ref{helicity_gas} shows an interesting additional effect of side chain size. Over the whole temperature range the radius of gyration of {\it Lys$_{10}$} is clearly larger than that of the other molecules. However, in the high temperature phase and in the transition regime the values of $r_{gy}$ for {\it Glu$_{10}$} and {\it Gln$_{10}$} as well as for {\it Asn$_{10}$} and {\it Asp$_{10}$} are very close to each other, despite differently sized side chains. Only in the low temperature  phase do their sizes become recognizably different (see also below the discussion of their respective low-energy structures).  Above this phase the size of these four polymers is apparently dominated by backbone fluctuations. The {\it Lys$_{10}$} side chains, however, being two or three CH$_2$ groups longer, respectively, do give a considerable contribution to the size of the molecule, as expected. 

\begin{table}
\caption{\label{tab:table3} Properties of the side chain ordering transitions observed for {\it Glu$_{10}$ } and {\it Gln$_{10}$ }: Transition temperature $T_2$, half width $\Delta T$, and specific heat at the transition temperature, $C(T_2)$. The number in parenthesis is the error in the last digit. }
\begin{ruledtabular}
\begin{tabular}{lrrrrrr}
 & & vacuum & & & solvent & \\
 & $T_2$[K] & $\Delta T$\footnotemark[1][K] & $C(T_2)$ [Kcal/mol] & 
   $T_2$[K] & $\Delta T$\footnotemark[1][K] & $C(T_2)$ [Kcal/mol] \\
\hline
{\it Glu$_{10}$ } &  166(16) & 203 &  0.176(4) & 111(10)  & 166 &  0.152(2)   \\
{\it Gln$_{10}$ } &  181(23) & 187 &  0.161(6) & 120(17)  & 146 &  0.133(3)    \\
\end{tabular}
\end{ruledtabular}
\footnotetext[1]{determined at $C=\left[C(T_2)+C(T_{min})\right]/2$, with $T_{min}$ from $C'(T_{min})=0$.}
\end{table}

Interestingly, a second peak in the specific heat is observed for two of the homopolymers, {\it Glu$_{10}$} and {\it Gln$_{10}$}. The peak temperatures $T_2$ are around $170K$, with a peak width of about $200K$. The exact peak properties are listed in Table~\ref{tab:table3}.
From the inlay of Fig.~\ref{helicity_gas} it follows that this second peak
cannot be related an additional collapse of the two molecules. We have argued in Ref.~\onlinecite{Wei2006} that this low-temperature peak is related to an ordering of the side chains in the form of hydrogen bonding between them. This hypothesis is supported by  Fig.~\ref{totalHB_gas} where we display the average total number $n_{hb}$ of hydrogen bonds as a function of temperature. Note that the fully formed helix contains six hydrogen bonds. Below $T=400K$ however, more than six hydrogen bonds are formed, with the number increasing with decreasing temperature. Since no hydrogen bond partners are available anymore on the backbone, these additional hydrogen bonds are formed between the side chains. 
The fluctuations of the hydrogen bonds, $\chi(T)=\left<\left(n_{hb}-\left<n_{hb}\right>\right)^2\right>$, are shown in the inlay of Fig.~\ref{totalHB_gas}. In addition to the peak at $T_1$ that appears for all polymers and that we expected from the helix coil transition, there is a second peak at the temperature $T_2$ in the curves for {\it Glu$_{10}$} and {\it Gln$_{10}$}.
This peak corresponds to the second peak in the specific heat, and this result clearly indicates indeed a second transition connected to the hydrogen bond formation. It separates  a low-temperature phase with ordered side chains from a phase at temperatures above $T_2$ where only the backbone is ordered and there is only a small number of fluctuating side chains. The form of the side chain ordering can be seen best from the lowest energy conformations displayed in Fig.~\ref{vac_structures}. Here, the side chains nestle along the cylinder formed by the helix and are stabilized by the side chain hydrogen bonds (not shown in the figure).  

{\it Asn$_{10}$} and {\it Asp$_{10}$}, whose side chains are only one CH$_2$-group shorter than those of {\it Glu$_{10}$} and {\it Gln$_{10}$}, show no such a second peak in the specific heat. This is surprising since also in these polymers hydrogen bonds are forming continuously with decreasing temperature among the side chains, like in {\it Glu$_{10}$} and {\it Gln$_{10}$} (see Fig.~\ref{totalHB_gas}). However, as the inlay of that figure demonstrates, this hydrogen bond formation is $not$ accompanied by a peak in the hydrogen bond number fluctuations.

We have investigated this puzzle by a more detailled inspection of the hydrogen bond formation. 
Fig.~\ref{sidechain_gas} shows the temperature dependence of the hydrogen bonds between monomers that are less than four units apart in the sequence, i.e. non-backbone hydrogen bonds. It can be seen that in {\it Asn$_{10}$} and {\it Asp$_{10}$} such bonds take part in the hydrogen bond formation and contribute to the fluctuations already at the helix-coil transition.  {\it Glu$_{10}$} and {\it Gln$_{10}$}, in contrast, show no such fluctuation peak. In addition, we observe that the number of these hydrogen bonds that are of short range in the sequence is $not $ sufficient to account for all hydrogen bonds in the low temperature phase. This means that {\it Asn$_{10}$} and {\it Asp$_{10}$} exhibit some side chain hydrogen bonds in the low temperature phase that span four residues and are, therefore, parallel to the backbone hydrogen bonds. In contrast, practically all side chain hydrogen bonds in {\it Glu$_{10}$} and {\it Gln$_{10}$} cover smaller distances along the sequence. Once the helix is formed, residues four units apart are actually closest in space. Consequently, side chain hydrogen bonds in {\it Asn$_{10}$} and {\it Asp$_{10}$} are actually between groups closer in space than they are in the case of {\it Glu$_{10}$} and {\it Gln$_{10}$}. This is also suggested when comparing the lowest energy structures of the four polymers in Fig.~\ref{sidechain_gas}. Although in all cases the side chains nestle along the cylinder formed by the helix, in {\it Asn$_{10}$} and {\it Asp$_{10}$} they appear to be more stretched and densely packed. In particular, in {\it Asn$_{10}$} and {\it Asp$_{10}$} the side chains appear to have 
much less possibility to find alternative hydrogen bond partners than they have in the case of {\it Glu$_{10}$} and {\it Gln$_{10}$}, leading to this dramatic difference in the fluctuation behavior of both molecule groups.

The reason for this remarkably different behavior of {\it Asn$_{10}$} and {\it Asp$_{10}$}  is the apparent existence  of a threshold length for the side chains. If these chains are not long enough, the number of degrees of freedom is simply too small to allow for the fluctuations observed for {\it Glu$_{10}$} and {\it Gln$_{10}$}. This difference can be compared  to the behavior of a generic model for phase transitions, the Ising model\cite{IsingRef}, in one and higher dimensions. While in dimension $d>1$ the number of effectively interacting degrees of freedom are sufficient to allow for a phase transition at some $T>0$, they are not sufficient in the 1d Ising model, leading only to a monotonic ordering upon decreasing the temperature to $T=0$ in that case.
Similarly, the ordering of the side chains in {\it Asn$_{10}$} and {\it Asp$_{10}$} does not have a transition like character but is spread out over a wide range of temperatures. 

Such a side chain ordering behavior is also what we observe for {\it Lys$_{10}$}. Since its side groups could participate only in weak hydrogen bonds, they act neutral at the helix coil transition and finally order themselves as it is shown in Fig.~\ref{vac_structures}. The side groups align themselves along the helical axis, however less densely packed than in the case of the strong hydrogen bond forming side chains. We observe a minor increase of the specific heat at the lower temperature end of our data. However, since the error in this regime is largest, we consider that finding inconclusive. Rather, we have to conclude that the specific heat actually stabilizes over a wide range of temperatures below the helix coil transition and side chain ordering is actually a smooth process over this temperature range.

We note that the structural features in the low temperature phase found here correspond to the results of theoretical structure investigations of such homopolymers in vacuum performed by H. Scheraga's group some 40 years ago\cite{Scheraga66,Scheraga67,Scheraga68,Scheraga70}.

So far, we have focused on the behavior of our molecules in gas phase.  This research is in itself theoretically interesting as well as important experimentally, since the properties of biopolymers in gas phase have become accessible to measurements only recently\cite{Jarrold2003b,Jarrold2004}. In particular it has been verified that helices can be stable in the gas phase up to high temperatures. The high transition temperatures that we observe in our simulations are, therefore, not unrealistic.
However, in nature proteins are usually solvated, and their function often depends on the details of the solvent environment. For this reason we have extended our investigation in a second step to that of solvated homopolymers.  

As in the case of gas phase simulations we observe for all five molecules a helix-coil transition characterized by peaks in the specific heat, see Fig.~\ref{C_sol}, their properties being listed also in Table~\ref{tab:table2}. As can be seen clearly, the width of these peaks is much narrower and their height is larger than in gas phase, indicating a sharper transition, a feature that has been found before for Polyalanine \cite{Hansmann1999,Peng2003} and by us for {\it Glu$_{10}$} \cite{Wei2006}. Correspondingly Fig.~\ref{helicity_sol} shows a rapid increase in the average helicity and a sharp drop of the radius of gyration (see the inlay of Fig.~\ref{helicity_sol}). 

All transition temperatures are  shifted considerably to lower values. The reason for this shift is the competition between the formation of backbone hydrogen bonds that stabilize an $\alpha$-helix, and the formation of hydrogen bonds between the backbone and the solvent in the coil phase,
the energy contribution of the latter being described in a mean field way by the solvent energy term (\ref{solventEnergy}). While in vacuum the
transition to the coil phase is driven solely by entropy, here also a part of the energy, i.e. the peptide-solvent interaction, favors the coil phase. These effects collaborate so that the transition takes place at a lower temperature and becomes sharper. However, the relative ordering of the transition temperatures among the polymers remains practically the same as in the gas phase, and the explanations for this ordering we gave above apply here, too.

Interestingly, again a second peak in the specific heat is observed for {\it Glu$_{10}$} and {\it Gln$_{10}$}, the exact properties being listed also in Table~\ref{tab:table3}. As before, this second peak
cannot be related to a collapse of the two molecules, as can be seen from the inlay of Fig.~\ref{helicity_sol}. However, here the reason for such a behavior cannot be the formation of hydrogen bonds among the side chains, since they interact with the solvent instead. The lowest energy structures show that the side chains extend into the solvent, see Fig.~\ref{structures_sol}. Also, Fig.~\ref{totalHB_sol} shows that the total number of hydrogen bonds is limited to those that stabilize the helix backbone.

A closer inspection shows that this second peak in the specific heat is mainly due to fluctuations in the solvent contribution to the total energy. This was already found in Ref.~\onlinecite{Wei2006} for {\it Glu$_{10}$}, and we confirm it here for {\it Gln$_{10}$}. The details of the individual contributions of the
fluctuations to generate the specific heat peak are rather subtle, however,
and we refer to that reference for a more detailed discussion. We just emphasize that - despite the fact that the side chain ground state structures are fundamentally different from gas phase - a side chain ordering transition is observed in solvent for {\it Glu$_{10}$} and {\it Gln$_{10}$}.

No such behavior can be observed for the shorter side chain polymers 
{\it Asn$_{10}$} and {\it Asp$_{10}$}. The lowest energy structures show that also in their case the side chains extend into the solvent, see Fig.~\ref{structures_sol}. However, the chains are apparently just too short to generate the necessary amount of solvent fluctuations. Again, as before in gas phase, we encounter the situation that the length of these side chains appears to be just below some threshold. Surprisingly there is still some minor side chain hydrogen bond formation in the low temperature phase of {\it Asn$_{10}$} and {\it Asp$_{10}$}, as can be seen in Fig.~\ref{totalHB_sol}. And as we saw in gas phase, in Fig.~\ref{sidechain_sol} we again observe enhanced side chain hydrogen bond fluctuations at the helix-coil transition, albeit on a smaller level. 

{\it Lys$_{10}$}, finally, exhibits again no particular features with respect to side chain ordering. 
As to be expected, its side groups extend far into the solvent, see Fig.~\ref{structures_sol}. As before in gas phase, this side chain ordering 
does not have a transition like character but is spread out over a wide range of temperatures.

\section{Summary and Outlook}

We have investigated backbone and side chain ordering in five homopolymers, {\it Glu$_{10}$, Gln$_{10}$, Asp$_{10}$, Asn$_{10}$, and Lys$_{10}$}. Those molecules  were selected to cover a wide variety of possible interactions between the side chains of the monomers. All homopolymers undergo helix-coil transitions and we were able to explain the ordering of their respective transition temperatures.

We found that peptides with $long$ side chains that are capable of hydrogen bonding, i.e. {\it Glu$_{10}$,} and {\it Gln$_{10}$}, exhibit a second transition at lower temperatures connected with side chain ordering. Remarkably, this  occurs in gas phase as well as in solvent, i.e. independent of the environment, although the character of the side chain structure is different in each case. However, $short$ side chains capable of hydrogen bonding, as in {\it Asp$_{10}$} and {\it Asn$_{10}$}, exhibit $no$ separate side chain ordering transition at temperatures below the helix-coil transition temperature. Instead,
final side chain ordering in these polymers takes place over a wide temperature range, exhibiting no phase transition like character. Side chain ordering in {\it  Lys$_{10}$}, whose side groups are long and polar, also takes place over a wide temperature range and exhibits no phase transition like character. Again, these results are independent of the environment, despite the different character of the side chain structure in each case.

 Our results indicate that, the de-coupling of backbone and side-chain ordering does not depend on the details of the environment, but solely depend on the particular side groups involved. While homopolymers with some side groups show a separate ordering transition in both environments, some do not.
Since natural proteins are heteropolymers of amino acids, there will be groups in a sequence that show separate side chain ordering in the homopolymer case as well as groups that do not. Whether  separate ordering still occurs in such a situation is unclear and will depend on the details of the sequence. However, it is an interesting question that should be tackled in the future.
Possible candidates for test proteins are the villin head piece subdomain HP-36 or the human parathyroid hormone fragment PTH(1-34) that have already been proven to be accessible by the computational methods used here \cite{HP36,PTH134}.

{\em Acknowledgments }
Support by a research grant (CHE-0313618) of the National Science Foundation (USA) is acknowledged.


%

%
%
\clearpage
{\huge Figure captions:}

\begin{description}
\item{Fig.~1:} Time series of the energy for one of the five homopolymers, {\it Glu$_{10}$}. Over the course of the simulation the system performs a random walk in energy. The number of {\it independent} visits of the low-energy region gives a measure for the quality of low temperature quantities. This number can be approximated by the number of tunnelings between preselected $E_{150K}$ and $E_{750K}$ values, -132.4 Kcal/mol and -42.6 Kcal/mol, as separately indicated by the two horizontal lines in the plot.
 
\item{Fig.~2:}  Specific heat $C(T)$ as function of temperature $T$  for  the five homopolymers in gas phase as obtained from a multicanonical simulation with $5\times 10^6$ sweeps ($10^7$ sweeps for {\it Glu$_{10}$}). 
Error bars are included and are mostly about the symbol size or less.

\item{Fig.~3:} Average number of helical residues, $<n_H>(T)$, as function of temperature $T$ for the five homopolymers in gas phase as obtained from a multicanonical simulation 
with $5\times 10^6$ sweeps ($10^7$ sweeps for {\it Glu$_{10}$}). 
The inlay shows the corresponding average radius of gyration $<r_{gy}>(T)$.
Error bars are included and are mostly about the symbol size or less.

\item{Fig.~4:} Average total number of hydrogen bonds, $<n_{hb}>(T)$, as function of temperature $T$ for the five homopolymers in gas phase as obtained from a  multicanonical simulation  
with $5\times 10^6$ sweeps ($10^7$ sweeps for {\it Glu$_{10}$}). 
The inlay shows the corresponding fluctuations $\chi (T)$  as function of temperature $T$.
Error bars are included and are mostly about the symbol size or less.

\item{Fig.~5:} Average number of hydrogen bonds between monomers that are less than four units apart in the sequence, $<n_{hb}^S>(T)$, as function of temperature $T$ for the five homopolymers in gas phase as obtained from a  multicanonical simulation with $5\times 10^6$ sweeps ($10^7$ sweeps for {\it Glu$_{10}$}). Note that this number is a lower limit to the number of non-backbone hydrogen bonds, see text. The inlay shows the corresponding fluctuations $\chi (T)$  as function of temperature $T$.
Error bars are included and are mostly about the symbol size or less.

\item{Fig.~6:} Lowest energy configuration of the five homopolymers in gas phase as obtained from a multicanonical simulation  
with $5\times 10^6$ sweeps ($10^7$ sweeps for {\it Glu$_{10}$})
and subsequent minimization. Each structure is shown in top and side view to give a better impression of the actual 3D structure.
The pictures have been drawn with Pymol\cite{pymol}.

\item{Fig.~7:} Specific heat $C(T)$ as function of temperature $T$ for the solvated  homopolymers  as obtained from a multicanonical simulation `with $5\times 10^6$ sweeps. 
The inlay shows the corresponding average radius of gyration $<r_{gy}>(T)$.
Error bars are included and are mostly about the symbol size or less.
 
\item{Fig.~8} Average number of helical residues, $<n_H>(T)$, as function of temperature $T$ for the five homopolymers in gas phase as obtained from a multicanonical simulation with $5\times 10^6$ sweeps. 
The inlay shows the corresponding  average radius of gyration $<r_{gy}>(T)$ .
Error bars are included and are mostly about the symbol size or less.

\item{Fig.~9:} Lowest energy configuration of the five solvated  homopolymers as obtained from a  multicanonical simulation with $5\times 10^6$ sweeps and subsequent minimization. Each structure is shown in top and side view to give a better impression of the actual 3D structure.
The pictures have been drawn with Pymol\cite{pymol}.

\item{Fig.~10:} Average total number of hydrogen bonds $<n_{hb}>(T)$ as function of temperature $T$ for the five homopolymers in gas phase as obtained from a  multicanonical simulation with $5\times 10^6$ sweeps. 
The inlay shows the corresponding fluctuation $\chi(T)$  as function of temperature $T$.
Error bars are included and are mostly about the symbol size or less.

\item{Fig.~11:} Average number of hydrogen bonds between monomers that are less than four units apart in the sequence, $<n_{hb}^S>(T)$, as function of temperature $T$ for the five homopolymers in gas phase as obtained from a  multicanonical simulation with $5\times 10^6$ sweeps. Note that this number is a lower limit to the number of non-backbone hydrogen bonds, see text. 
The inlay shows the corresponding fluctuation $\chi(T)$  as function of temperature $T$.
Error bars are included and are mostly about the symbol size or less.

\item{Fig.~12:} TOC Graphic

\end{description}
%


%
\setcounter{figure}{0}
%
\clearpage
\begin{sidewaysfigure}
    \includegraphics[width=1.0\columnwidth]{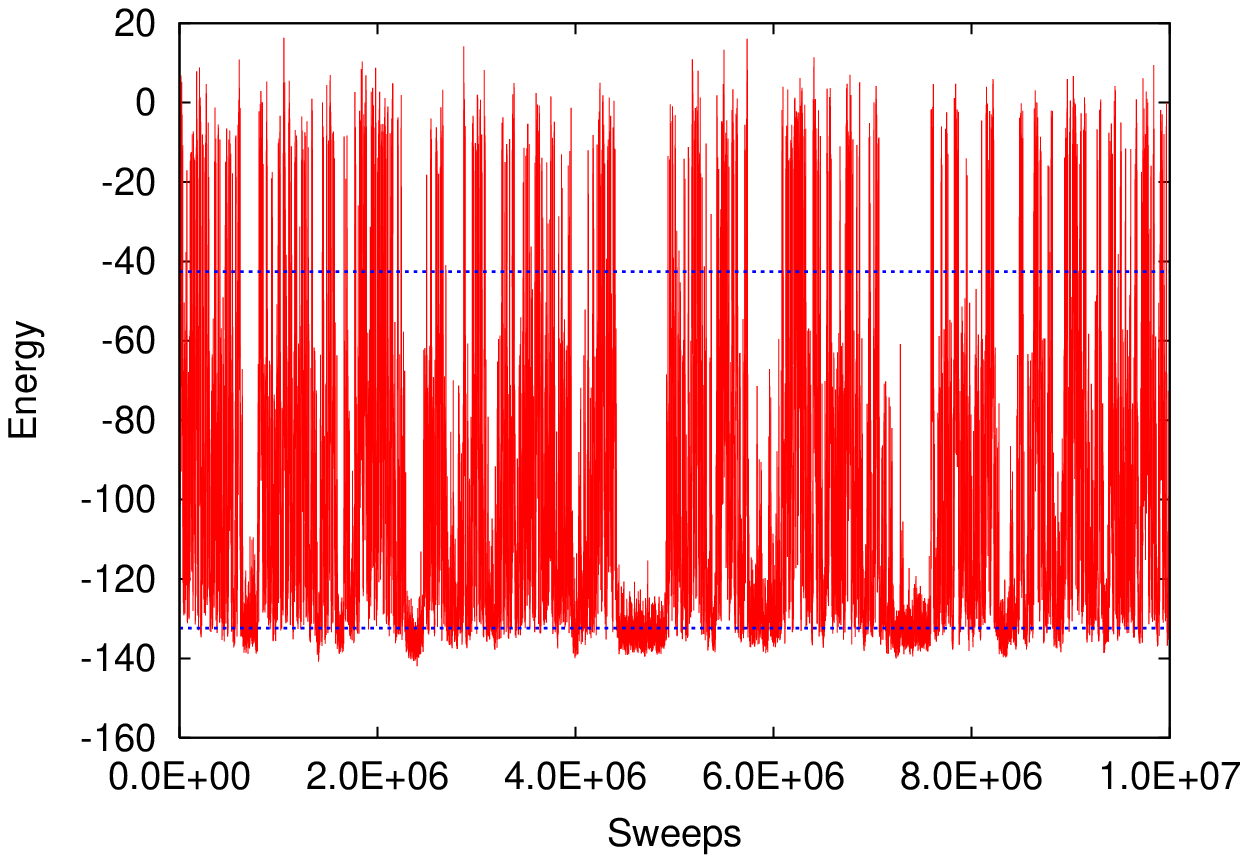}
\caption{\label{glu_vac}}
\end{sidewaysfigure}
%
\clearpage
\begin{sidewaysfigure}
    \includegraphics[width=1.0\columnwidth]{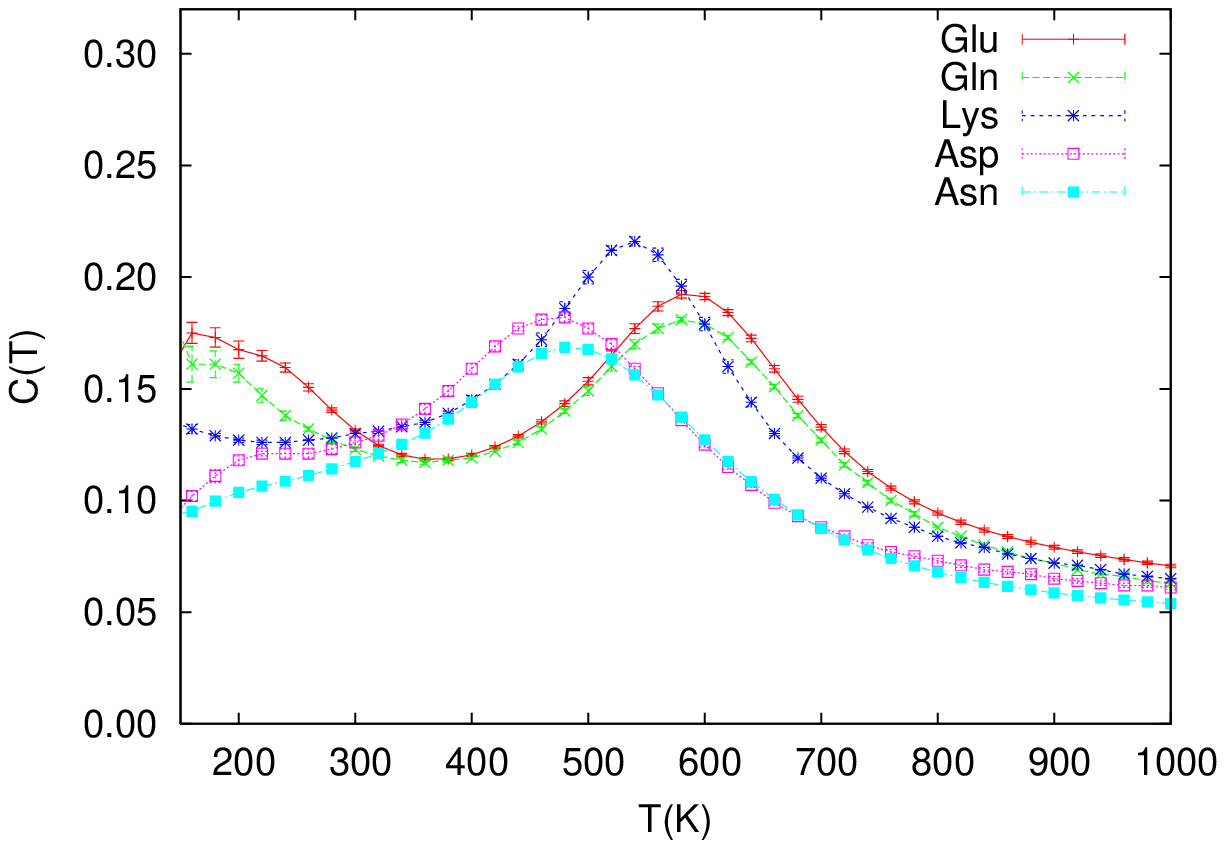}
\caption{\label{C_gas}}
\end{sidewaysfigure}
%
\clearpage
\begin{sidewaysfigure}
  \includegraphics[width=1.0\columnwidth]{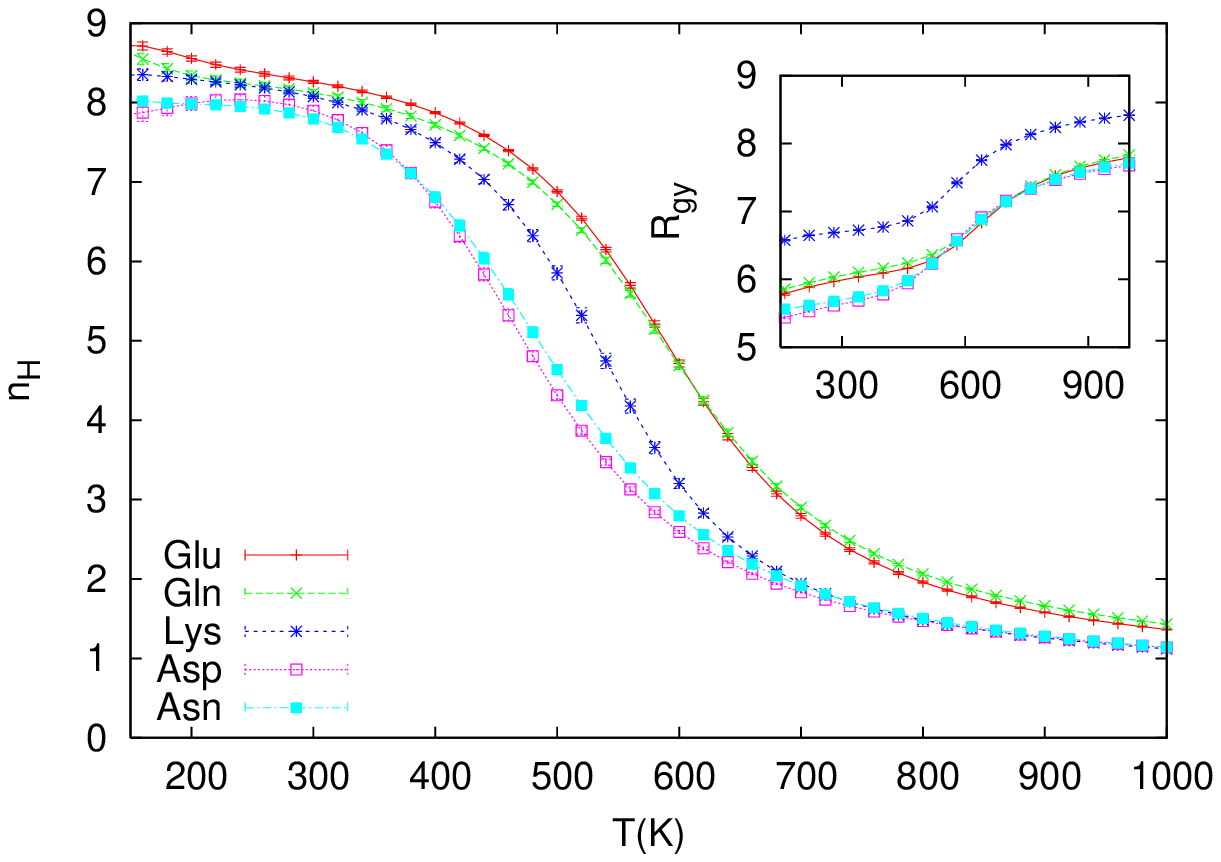}
\caption{\label{helicity_gas}}
\end{sidewaysfigure}
%
\clearpage
\begin{sidewaysfigure}
   \includegraphics[width=1.0\columnwidth]{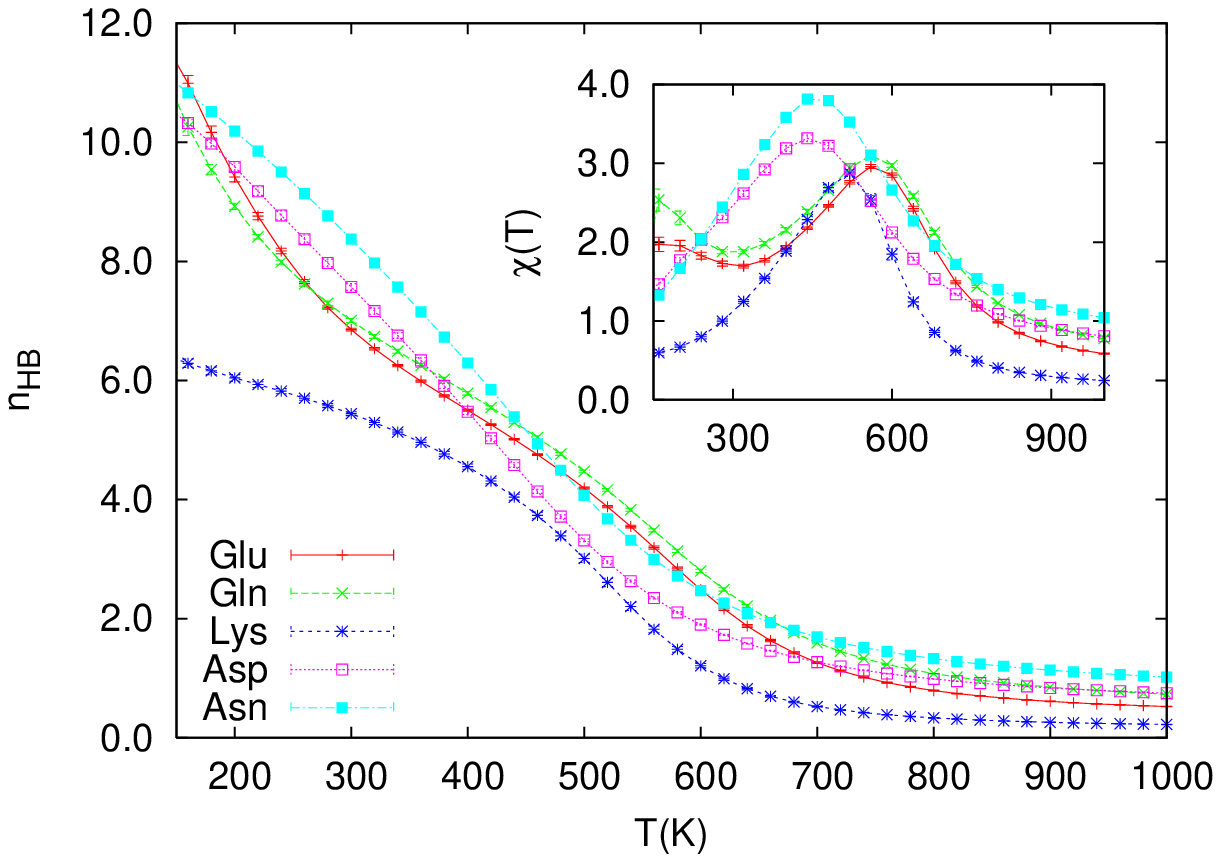}
\caption{\label{totalHB_gas}}
\end{sidewaysfigure}
%
\clearpage
\begin{sidewaysfigure}
   \includegraphics[width=1.0\columnwidth]{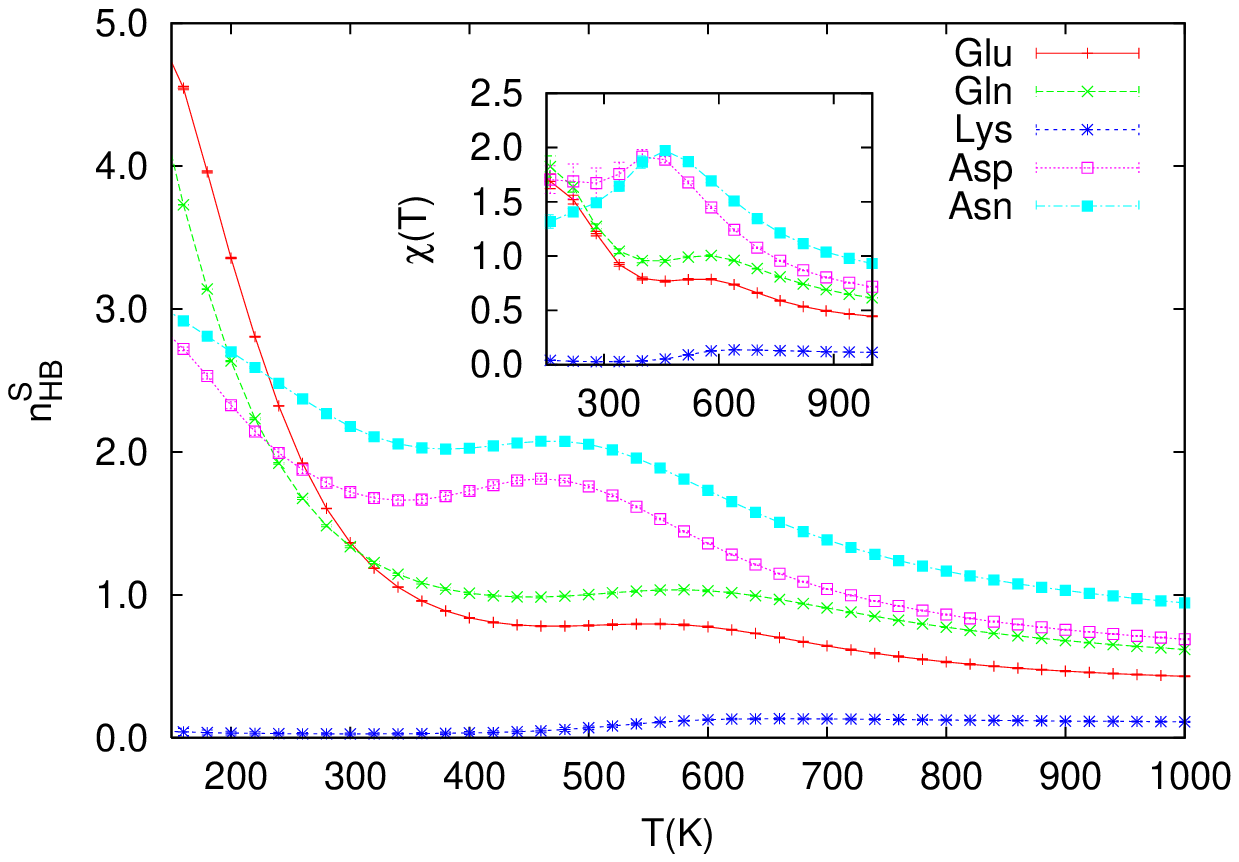}
\caption{\label{sidechain_gas}}
\end{sidewaysfigure}
%
\clearpage
\begin{figure}
   \includegraphics[width=1.0\columnwidth]{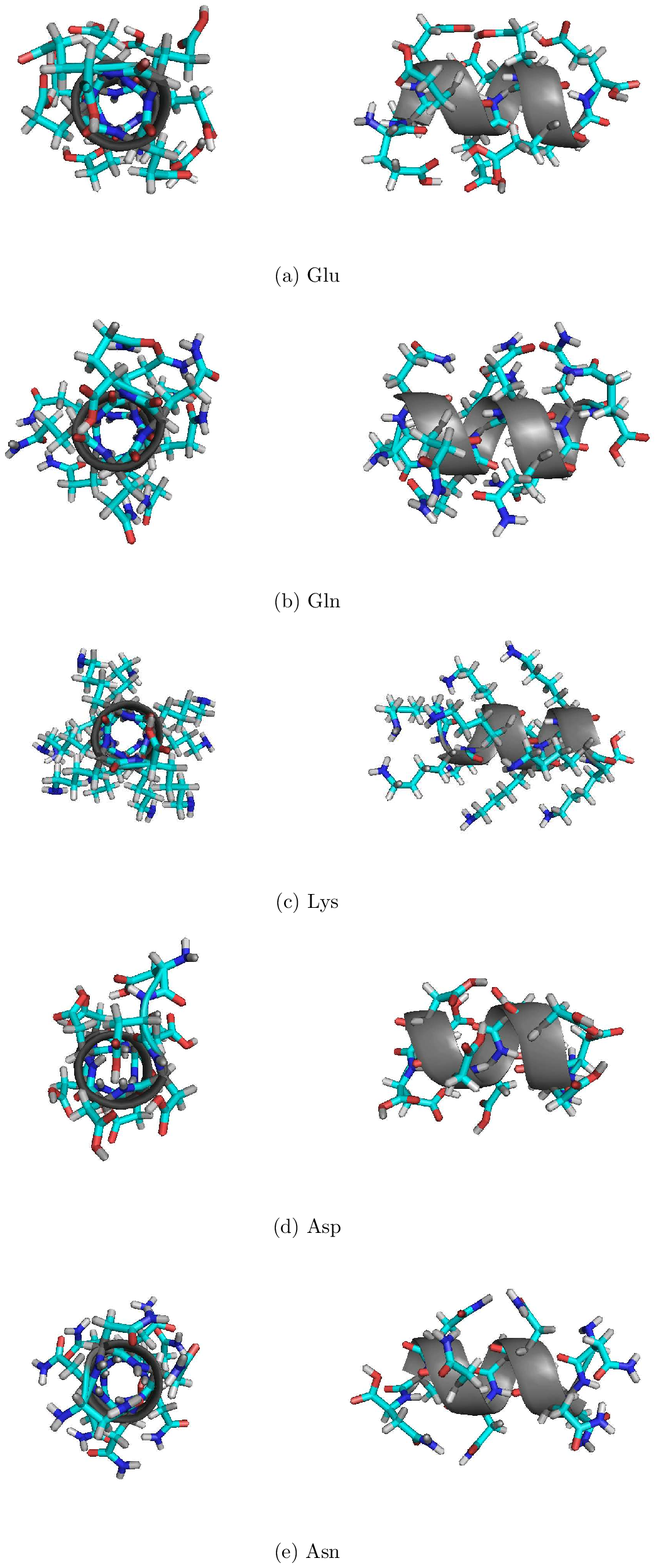}
\caption{\label{vac_structures}}
\end{figure}
%
\clearpage
\begin{sidewaysfigure}
   \includegraphics[width=1.0\columnwidth]{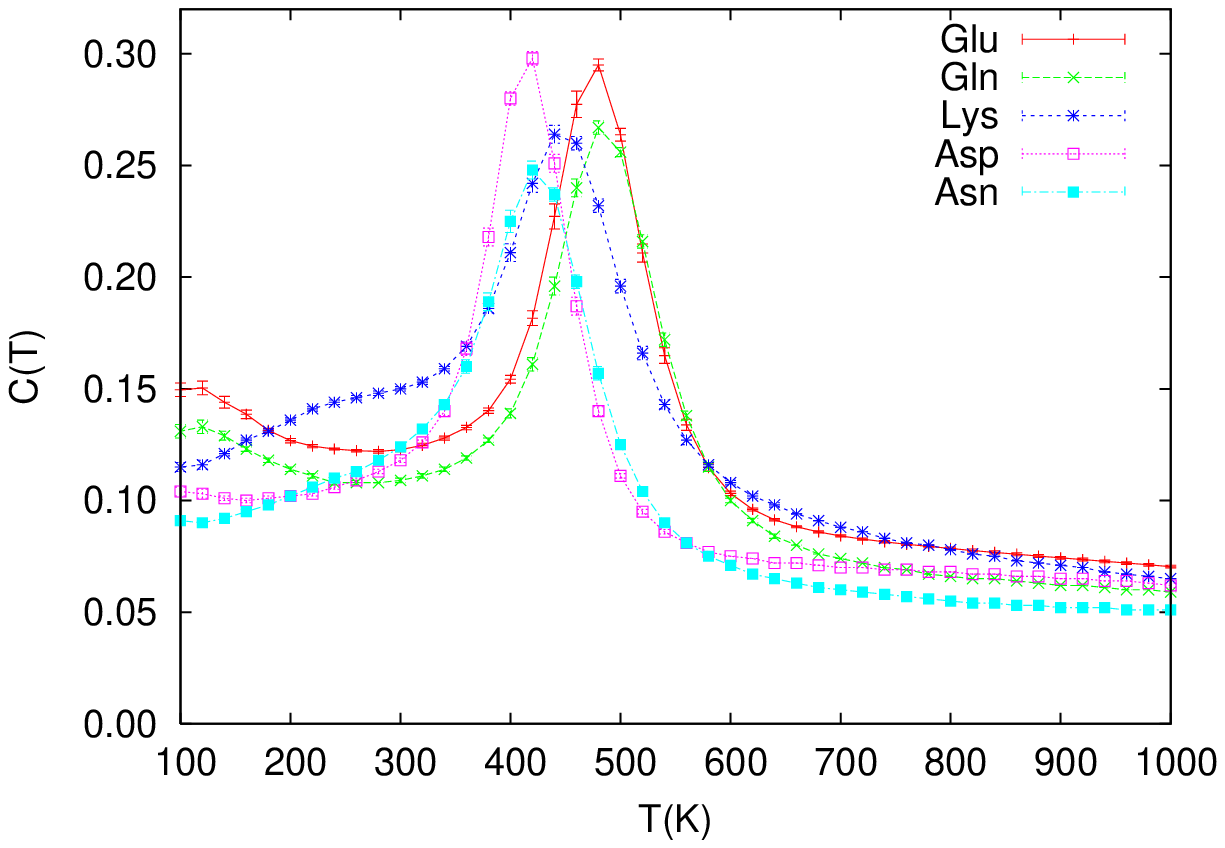}
\caption{\label{C_sol}}
\end{sidewaysfigure}
\clearpage
\begin{sidewaysfigure}
   \includegraphics[width=1.0\columnwidth]{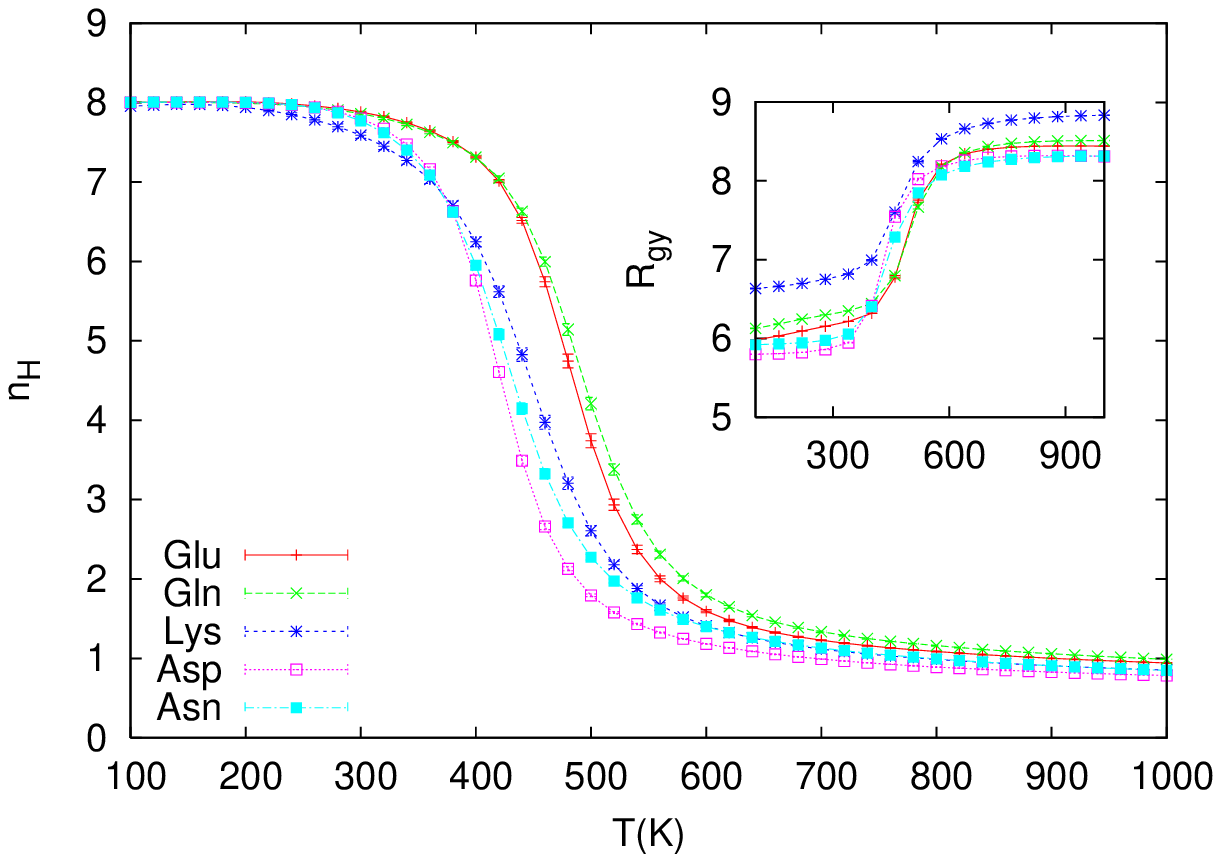}
\caption{\label{helicity_sol}}
\end{sidewaysfigure}
%
\clearpage
\begin{figure}
   \includegraphics[width=1.0\columnwidth]{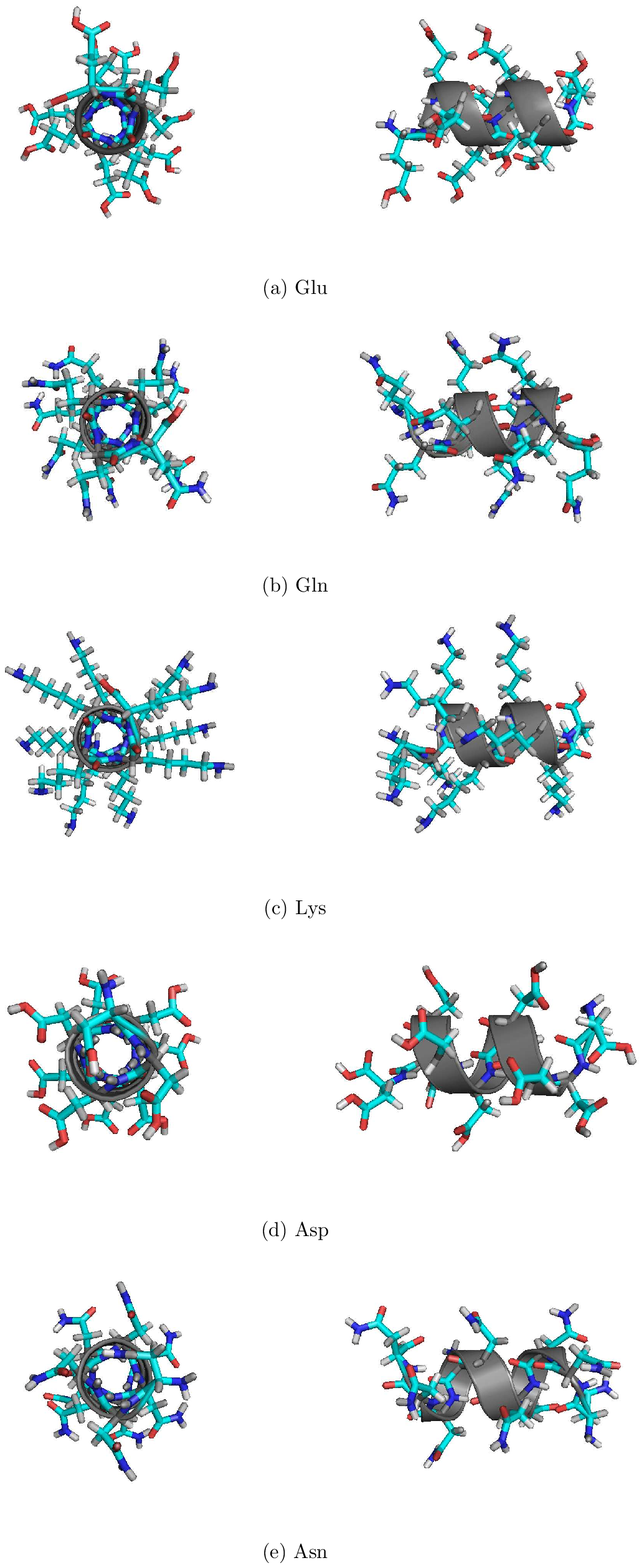}
\caption{\label{structures_sol}}
\end{figure}
%
\clearpage
\begin{sidewaysfigure}
   \includegraphics[width=1.0\columnwidth]{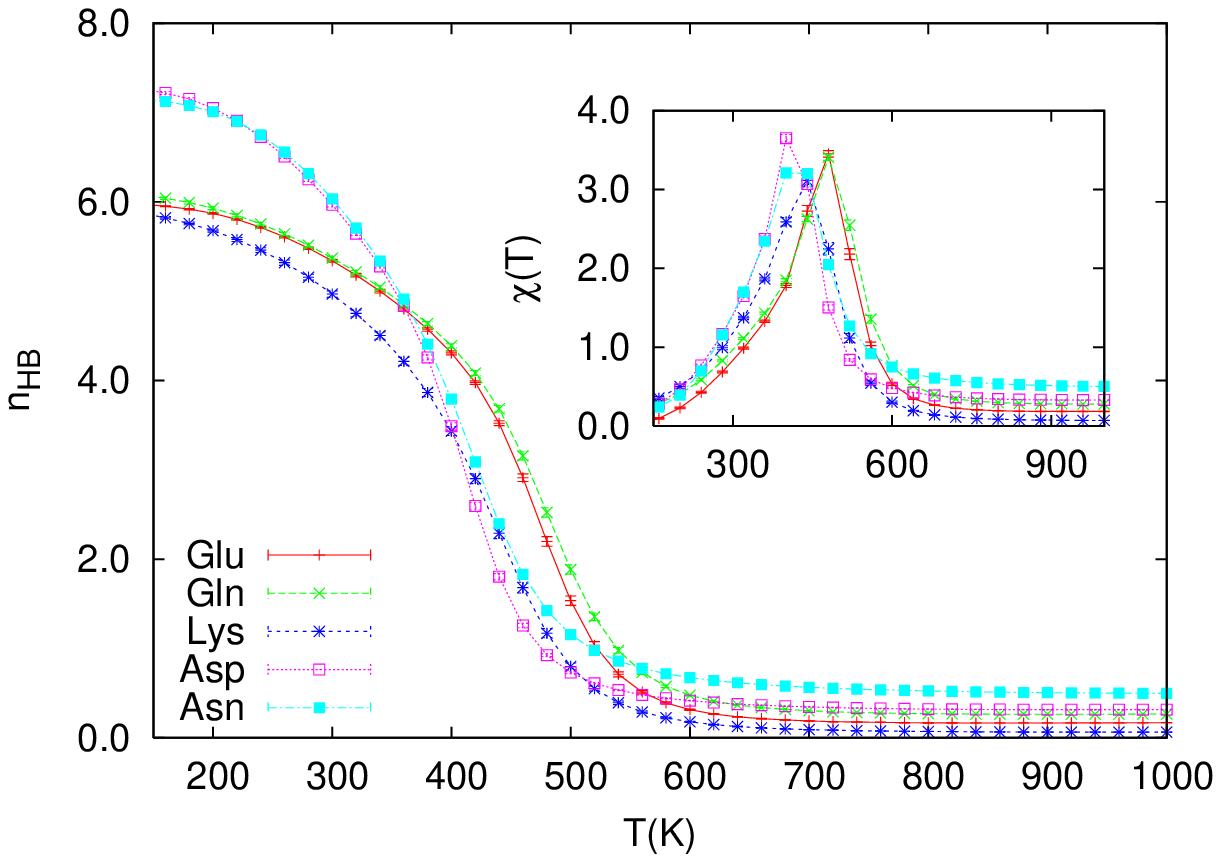}
\caption{\label{totalHB_sol}}
\end{sidewaysfigure}
%
\clearpage
\begin{sidewaysfigure}
   \includegraphics[width=1.0\columnwidth]{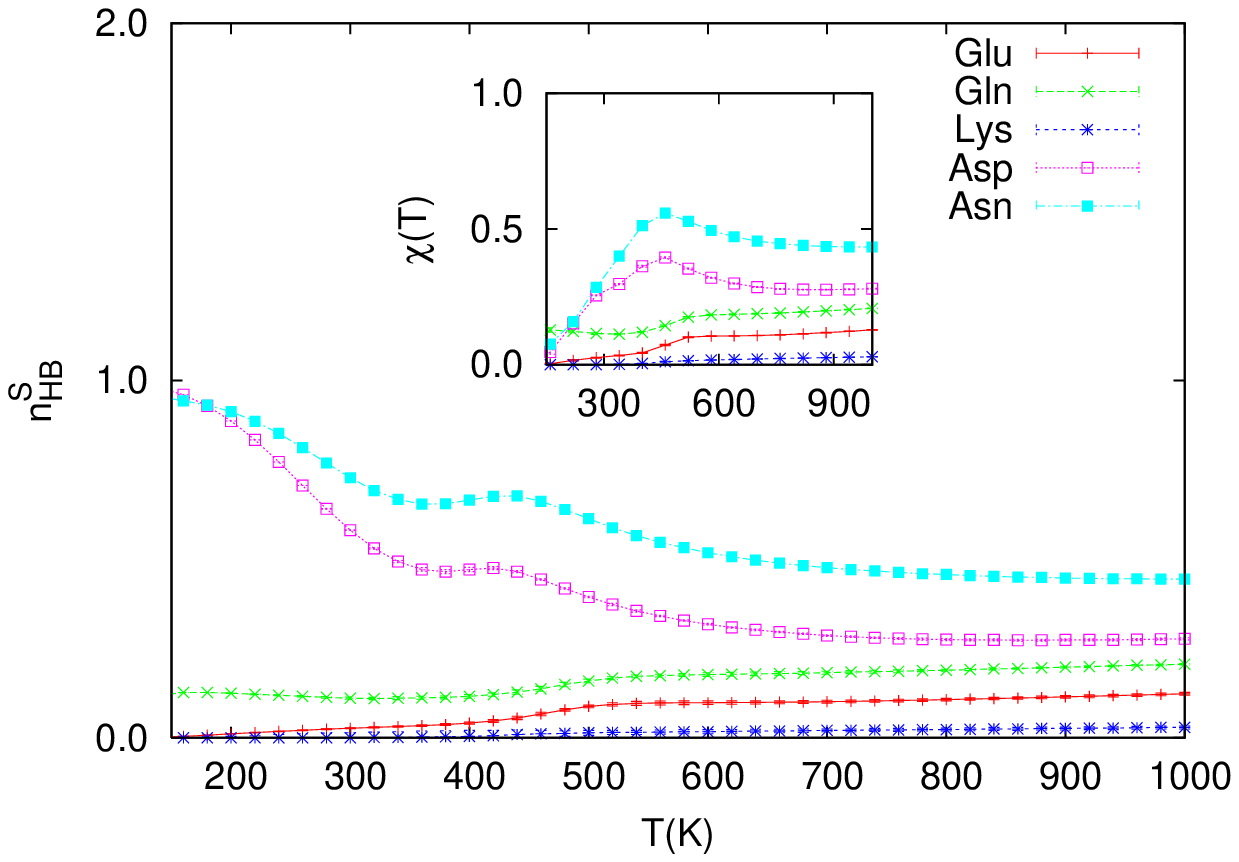}
\caption{\label{sidechain_sol}}
\end{sidewaysfigure}
%
\clearpage
\begin{sidewaysfigure}
   \includegraphics{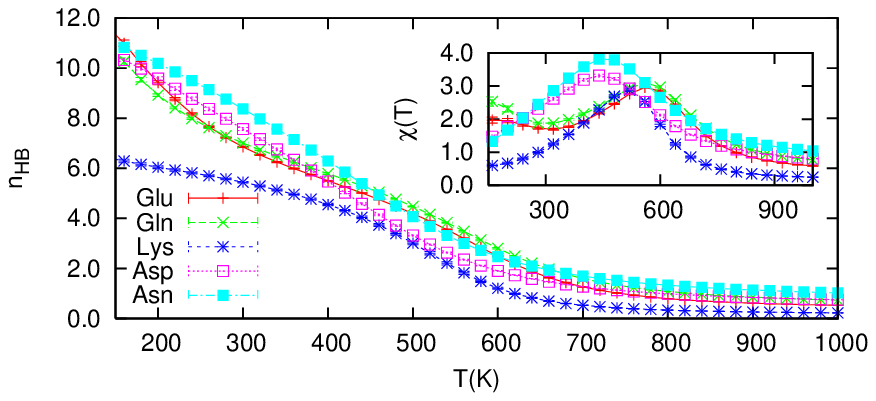}
\caption{\label{TOC}}
\end{sidewaysfigure}


\begin{thebibliography}{99}
\bibitem{sidechainHBFootnote} 
We note that this criterion was sufficient in \onlinecite{Wei2006} to distinguish between side chain and backbone hydrogen bonds. We will see that this is not the case here.
\end{thebibliography}
\end{document}